\documentclass[conference]{IEEEtran}
\usepackage{amssymb}
\usepackage{color}
\usepackage{multirow}
\usepackage{mathrsfs}

\newtheorem{definition}{Definition}

\usepackage{cite}
\ifCLASSINFOpdf
  \usepackage[pdftex]{graphicx}
  \graphicspath{{../pdf/}{../jpeg/}}
  \DeclareGraphicsExtensions{.pdf,.jpeg,.png}
\else
  \usepackage[dvips]{graphicx}
  \graphicspath{{../eps/}}
  \DeclareGraphicsExtensions{.eps}
\fi
\usepackage[cmex10]{amsmath}
\usepackage{array}
\usepackage[tight,footnotesize]{subfigure}
\usepackage{stfloats}

\usepackage{arydshln}
\usepackage{rotating}
\usepackage{microtype}
\def\gap{.5ex}
\abovedisplayskip\gap
\belowdisplayskip\gap
\abovedisplayshortskip\gap
\belowdisplayshortskip\gap

\IEEEoverridecommandlockouts

\usepackage{epstopdf}
\begin{document}
\title{Centralized Caching with Unequal Cache Sizes}

\author{\IEEEauthorblockN{Behzad Asadi, Lawrence Ong, and Sarah J. Johnson}
	\IEEEauthorblockA{School of Electrical Engineering and Computing, The University of Newcastle, Australia}
	Email: {behzad.asadi@uon.edu.au, lawrence.ong@newcastle.edu.au, sarah.johnson@newcastle.edu.au}
}

\maketitle

\begin{abstract}
We address a centralized caching problem with unequal cache sizes. We consider a system with a server of files connected through a shared error-free link to a group of cache-enabled users where one subgroup has a larger cache size than the other. We propose an explicit caching scheme for the considered system aimed at minimizing the load of worst-case demands over the shared link. As suggested by numerical evaluations, our scheme improves upon the best existing explicit scheme by having a lower worst-case load; also, our scheme performs within a multiplicative factor of 1.11 from the scheme that can be obtained by solving an optimisation problem in which the number of parameters grows exponentially with the number of users.
\end{abstract}
\IEEEpeerreviewmaketitle

\section{Introduction}\label{Sec:Introduction}
Content traffic, which is the dominant form of traffic in data communication networks, is not uniformly distributed over the day. This makes caching an integral part of data networks in order to tackle the non-uniformity of traffic. Caching schemes consist of two phases for content delivery. In the first phase, called the placement phase, content is partly placed in caches close to users. This phase takes place during off-peak hours when the requests of users are still unknown. In the second phase, called the delivery phase, each user requests a file while having access to a cache of pre-fetched content. This phase takes place during peak hours when we need to minimize the load over the network.

The information-theoretic study of a network of caches originated with the work of Maddah-Ali and Niesen~\cite{CentralizedCaching}. They considered a centralized multicast set-up where there is a server of files connected via a shared error-free link to a group of users, each equipped with a dedicated cache of equal size. They introduced a caching gain called global caching gain. This gain is in addition to local caching gain, which is the result of the fact that users have access to part of their requested files. Global caching gain is achieved by simultaneously sending data to multiple users in the delivery phase via coded transmission over the shared link.

The information-theoretic study of cache-aided networks has then been extended to address other scenarios which arise in practice such as decentralized caching \cite{DecentralizedCaching}, where the identity or the number of users is not clear in the placement phase; caching with non-uniform file popularity \cite{CachingNonuniformDemands}, where some of the files in the server are more popular than the others; and hierarchical caching~\cite{HierarchicalCodedCaching}, where there are multiple layers of caches. Also, while most of existing works consider uncoded cache placement, where the cache of each user is populated by directly placing parts of the server files, it has been shown for some special cases that coded cache placement can outperform uncoded cache placement~\cite{CentralizedCaching, CachingWithCodedPlacement1, CachingWithCodedPlacement2, CachingWithCodedPlacement3}.

%-------------------------------------------------------------------------
\begin{figure}[t]
	\centering
	\includegraphics[width=0.46\textwidth]{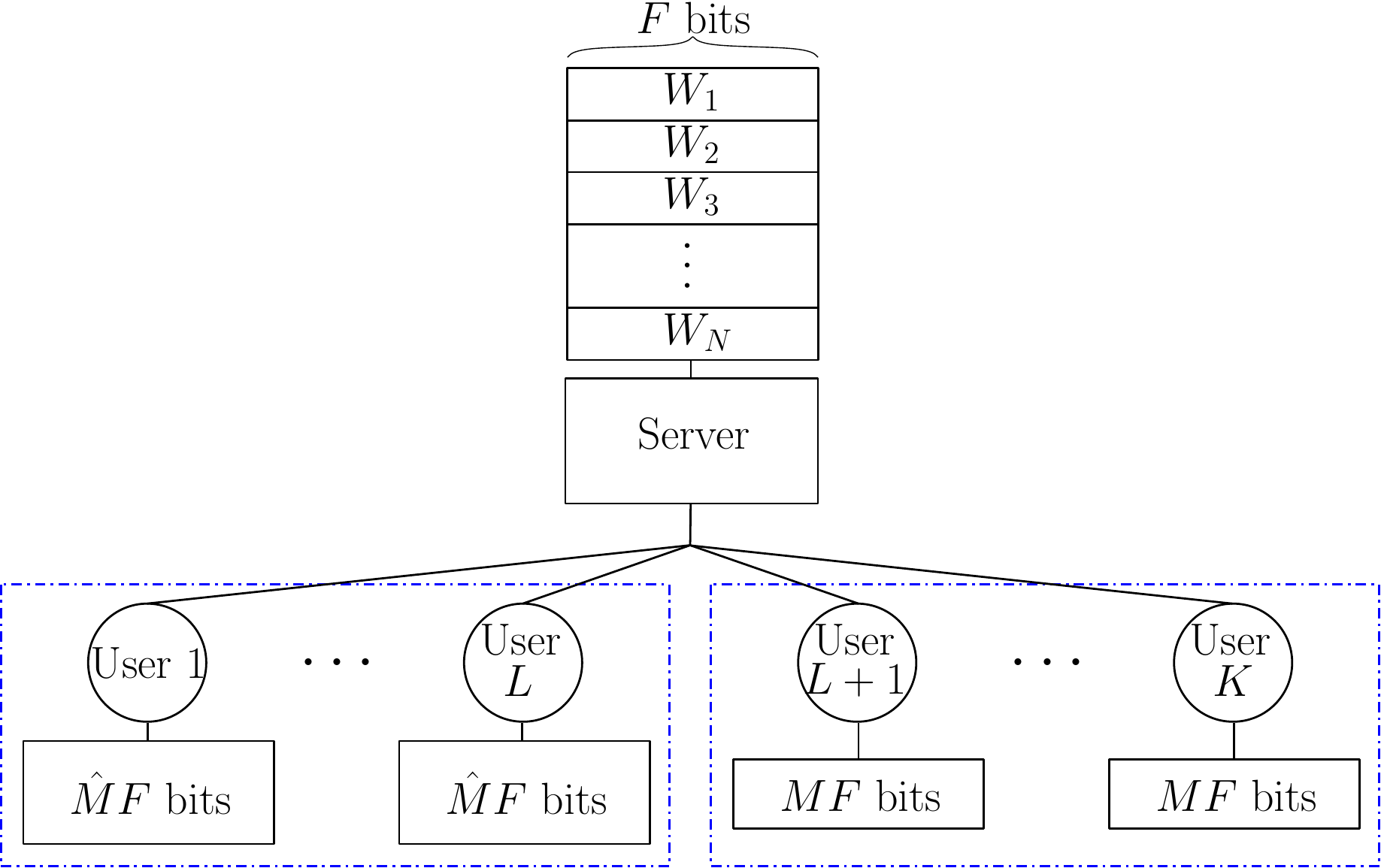}
	\vskip-10pt
	\caption{System model with a server storing $N$ files of size $F$ bits connected through a shared error-free link to $K$ users. User~$i$ is equipped with a cache of size $M_iF$ bits where $M_i=\hat{M}$, $1\leq i\leq L$, and $M_i=M$, $L+1\leq i\leq K$, for some $\hat{M}>M$.}
	\label{Fig:SystemModel}
	\vskip-15pt
\end{figure}
%-------------------------------------------------------------------------

\subsection{Existing works and Contributions}\label{Sec:ExistingWorksandContributions}
In this work, we address caching problems where there is a server connected through a shared error-free link to a group of users with caches of possibly different sizes. The objective is to minimize the load of worst-case demands over the shared link. Considering decentralized caching with unequal cache sizes, the placement phase is the same as the one for the equal-cache case where randomly part of each file is assigned to the cache of each user. The main challenge is to exploit all the coding opportunities in the delivery phase~\cite{DecentralizedUnequalCache1,DecentralizedUnequalCache2}. 

However, considering centralized caching with unequal cache sizes, the challenge also involves designing the placement phase. For the two-user case, Cao et al.~\cite{CentralizedUnequalCache3} proposed an optimum caching scheme, and showed that coded cache placement outperforms uncoded. For a system with an arbitrary number of users, Saeedi Bidokhti et al.~\cite{CentralizedUnequalCache1} proposed a scheme with uncoded cache placement constructed based on the memory sharing of the scheme for centralized caching with equal cache sizes~\cite{CentralizedCaching}. Also, Ibrahim et al.~\cite{CentralizedUnequalCache2}, assuming uncoded cache placement and linear coded delivery, formulated this problem  as a linear optimisation problem in which the number of parameters grows exponentially with the number of users. As the number of users grows, the scheme by Saeedi Bidokhti et al.~\cite{CentralizedUnequalCache1} remains simple at the cost of performance, and the optimisation problem by Ibrahim et al.~\cite{CentralizedUnequalCache2} becomes intractable.

In the light of the above mentioned issues, we propose a new caching scheme with uncoded cache placement for centralized caching with unequal cache sizes where there are two subgroups of users, one with a larger cache size than the other. Our caching scheme outperforms the caching scheme proposed by Saeedi Bidokhti et al.~\cite{CentralizedUnequalCache1} suggested by numerical evaluations. In comparison to the work by Ibrahim et al.~\cite{CentralizedUnequalCache2}, as our scheme is an explicit scheme, it does not have the complexity issue associated with solving an optimisation problem. Also, our scheme performs within a multiplicative factor of 1.11 from the scheme by Ibrahim et al.~\cite{CentralizedUnequalCache2} suggested by numerical evaluations.

\section{System Model}\label{Section:SystemModel}
We consider a centralized caching problem where there is a server storing $N$ independent 
files $W_\ell$, $\ell\in\mathcal{N}$, $\mathcal{N}=\{1,2,\ldots,N\}$, connected through a shared error-free link to $K$ cache-enabled users, as shown in Fig.~\ref{Fig:SystemModel}. We assume that the number of files in the server is at least as many as the number of users, i.e., $N\geq K$. Each file in the server is of size $F\in\mathbb{N}$ bits (where $\mathbb{N}$ is the set of natural numbers), and is uniformly distributed over the set $\mathcal{W}=\left\{1,2,\ldots,2^{F}\right\}$. User~$i$, $i\in\mathcal{K}$, $\mathcal{K}=\{1,2,\ldots,K\}$, is equipped with a cache of size $M_iF$ bits for some $M_i\in\mathbb{R}$, $0\leq M_i\leq N$, where $\mathbb{R}$ is the set of real numbers. The content of the cache of user~$i$ is denoted by $Z_i$. We represent all the cache sizes by the vector $\mathbf{M}=(M_1,M_2,\ldots,M_K)$. In this work, we assume that there are two subgroups of users, one with a larger cache size than the other, i.e., $M_i=\hat{M}$, $1\leq i \leq L$, and $M_i={M}$, $L+1\leq i \leq K$, for some $\hat{M}>M$. User~$i$ requests $W_{d_i}$ from the server where $d_i\in\mathcal{N}$. We represent the request of all the users by the vector $\mathbf{d}=(d_1,d_2,\ldots,d_K)$. User~$i$ needs to decode $W_{d_i}$ using $Z_i$, and the signal $X_\mathbf{d}$ transmitted by the server over the shared link.

As mentioned earlier, each caching scheme consists of two phases, the placement phase and the delivery phase. The placement phase consists of $K$ caching functions
\begin{align*}
\phi_i:\mathcal{W}^{N}\rightarrow \mathcal{Z}_i,\;\; i\in\mathcal{K},
\end{align*}
where $\mathcal{Z}_i\hskip-2pt=\hskip-2pt\left\{\hskip-2pt 1,2,\ldots,2^{\left\lfloor M_iF \right\rfloor}\hskip-2pt\right\}$, i.e., $Z_i\hskip-2pt=\hskip-2pt\phi_i\left(\hskip-2pt W_1,W_2,\ldots,W_N\hskip-2pt\right)$.

The delivery phase consists of $N^K$ encoding functions
\begin{align*}
\psi_{\mathbf{d}}:\mathcal{W}^{N}\rightarrow \mathcal{X},
\end{align*}
where $\mathcal{X}=\left\{1,2,\ldots,2^{\left\lfloor RF \right\rfloor}\right\}$, i.e.,
\begin{align*}
X_{\mathbf{d}}=\psi_{\mathbf{d}}\left(W_1,W_2,\ldots,W_N\right).
\end{align*}
We refer to $RF$ as the load of the transmission and $R$ as the rate of the transmission over the shared link.

The delivery phase consists of also $KN^K$ decoding functions
\begin{align*}
\theta_{\mathbf{d},i}:
\mathcal{Z}_i\times\mathcal{X}\rightarrow \mathcal{W},\;\;i\in\mathcal{K},
\end{align*}
i.e., $\hat{W}_{\mathbf{d},i}=\theta_{\mathbf{d},i}(X_{\mathbf{d}},Z_i)$, where $\hat{W}_{\mathbf{d},i}$ is the decoded version of $W_{d_i}$ at user~$i$ when the demand vector is $\mathbf{d}$.

The probability of error for the scheme is defined as
\begin{align*}
\underset{\mathbf{d}}{\max}\;\,\underset{i}{\max}\;P(\hat{W}_{\mathbf{d},i}\neq W_{d_i}).
\end{align*}

\begin{definition}
For a given $\mathbf{M}$, we say that the rate $R$ is achievable if for every $\epsilon>0$ and large enough $F$, there exists a caching scheme with rate $R$ such that its probability of error is less than $\epsilon$. For a given $\mathbf{M}$, we also define $R^{\star}(\mathbf{M})$ as the infimum of all achievable rates.
\end{definition}

\section{Background}\label{Sec:Background}
In this section, we first consider centralized caching with equal cache sizes, i.e., $M_i=M,\,\forall i$, and review the optimum scheme among those with uncoded placement~\cite{CentralizedCaching, OptimumCachingWithUnCodedPlacement}.  We then review existing works on centralized caching with unequal cache sizes where there are more than two users~\cite{CentralizedUnequalCache1,CentralizedUnequalCache2}.

\subsection{Equal Cache Sizes}\label{Sec:EqualCache}
Here, we present the optimum caching scheme for centralized caching with equal cache sizes when the cache placement is uncoded, and $N\geq K$~\cite{CentralizedCaching}. In this scheme, a parameter denoted by $t$ is defined at the beginning as
\begin{align*}
t=\frac{KM}{N}.
\end{align*}

First, assume that $t$ is an integer. As $0\leq M\leq N$, we have $t\in\{0,1,2,\ldots,K\}$. In the placement phase, $W_\ell$, $\ell\in\mathcal{N}$, is divided into $\binom{K}{t}$ non-overlapping parts denoted by $W_{\ell,\mathcal{T}}$ where $\mathcal{T}\subseteq\mathcal{K}$ and $\left|\mathcal{T}\right|=t$ ($\left|\mathcal{T}\right|$ denotes the cardinality of the set $\mathcal{T}$). $W_{\ell,\mathcal{T}}$ is then placed in the cache of user $i$ if $i\in\mathcal{T}$. This means that the size of each part is $\frac{F}{\binom{k}{t}}$ bits, and we place $\binom{K-1}{t-1}$ parts from each file in the cache of user~$i$. Therefore, we satisfy the cache size constraint as we have
\begin{align*}
N\frac{\binom{K-1}{t-1}}{\binom{K}{t}}=M.
\end{align*}
In the delivery phase, the server transmits 
\begin{align*}
X_{\mathbf{d},\mathcal{S}}=\underset{s\in\mathcal{S}}{\bigoplus} W_{d_s,\mathcal{S}\setminus s},
\end{align*}
for every $\mathcal{S}\subseteq\mathcal{K}$ where $\left|\mathcal{S}\right|=t+1$. This results in the transmission rate of
\begin{align*}
R_{\text{eq}}(N,K,M)=\frac{\binom{K}{t+1}}{\binom{K}{t}}.
\end{align*}
This delivery scheme satisfies the demands of all the $K$ users~\cite{CentralizedCaching}. 

Now, assume that $t$ is not an integer. In this case, memory sharing is utilized where $t_\text{int}$ is defined as
\begin{align*}
t_\text{int}\triangleq\left\lfloor t \right\rfloor,
\end{align*}
and $\alpha$ is computed using the following equation
\begin{align*}
M=\frac{tN}{K}=\alpha\frac{t_\text{int}N}{K}+(1-\alpha)\frac{(t_\text{int}+1)N}{K},
\end{align*}
where $0<\alpha\leq1$. Based on $\alpha$, the caching problem is divided into two independent problems. In the first one, the cache size is $\alpha\frac{t_\text{int}N}{K}F$, and we cache the first $\alpha F$ bits of the files, denoted by $W^{(\alpha)}_{\ell}$, $\ell\in\mathcal{N}$. In the delivery phase, the server transmits 
\begin{align}\label{Eq:Component1}
X^{(\alpha)}_{\mathbf{d},\mathcal{S}_1}=\underset{s\in\mathcal{S}_1}{\bigoplus} W^{(\alpha)}_{d_s,\mathcal{S}_1\setminus s},
\end{align}
for every $\mathcal{S}_1\subseteq\mathcal{K}$ where $\left|\mathcal{S}_1\right|=t_\text{int}+1$.

In the second one, the cache size is $(1-\alpha)\frac{(t_\text{int}+1)N}{K}F$, and we cache the last $(1-\alpha)F$ bits of the files, denoted by $W^{(1-\alpha)}_{\ell}$, $\ell\in\mathcal{N}$. In the delivery phase, the server transmits
\begin{align}\label{Eq:Component2}
X^{(1-\alpha)}_{\mathbf{d},\mathcal{S}_2}=\underset{s\in\mathcal{S}_2}{\bigoplus} W^{(1-\alpha)}_{d_s,\mathcal{S}_2\setminus s},
\end{align}
for every $\mathcal{S}_2\subseteq\mathcal{K}$ where $\left|\mathcal{S}_2\right|=t_\text{int}+2$.

Consequently, the rate
\begin{align}\label{Eq:Rate}
R_{\text{eq}}(N,K,M)=\alpha \frac{\binom{K}{t_\text{int}+1}}{\binom{K}{t_\text{int}}}+(1-\alpha)\frac{\binom{K}{t_\text{int}+2}}{\binom{K}{t_\text{int}+1}},
\end{align}
is achieved where $\binom{a}{b}$ is considered to be zero if $b>a$.

%-------------------------------------------------------------------------
\begin{figure}[t]
	\centering
	\includegraphics[width=0.3\textwidth]{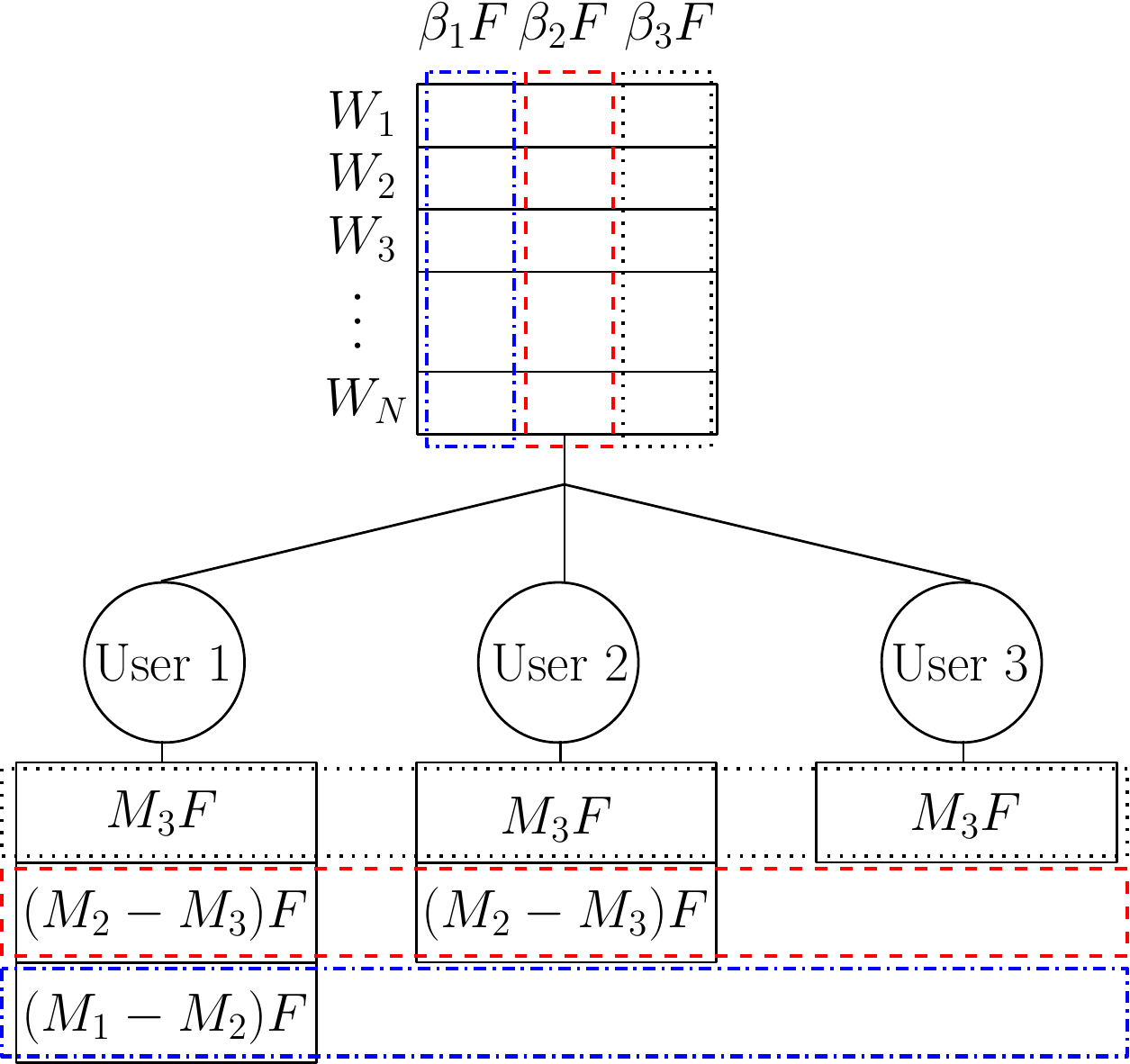}
	\vskip-10pt
	\caption{An existing scheme for centralized caching with unequal cache sizes} 
	\vskip-15pt
	\label{Fig:ExScheme1}
\end{figure}
%-------------------------------------------------------------------------

\subsection{Unequal Cache Sizes}\label{Sec:ExistingWorks}
Here, we present existing works on centralized caching with unequal cache sizes where there are more than two users.
\subsubsection{Scheme~1~\cite{CentralizedUnequalCache1}}\label{Sec:ExistingScheme1} In this scheme, assuming without loss of generality that $M_1\geq M_2 \geq \cdots \geq M_K$, the problem is divided into $K$ caching problems. In problem $i$, $i\in\mathcal{K}$, there are two groups of users: the first group is composed of users 1 to $i$, all with equal cache size of $(M_i-M_{i+1})F$ bits; the second group is composed of users $i+1$ to $K$, all without cache. In problem $K$, $M_{K+1}$ is considered as zero, and there is only one group consisting of $K$ users all with equal cache size of $M_KF$ bits. In problem $i$, we only consider $\beta_iF$ bits of the files where $\beta_1+\beta_2+\cdots+\beta_K=1$. This scheme is schematically shown in Fig.~\ref{Fig:ExScheme1} for the three-user case. Based on the equal cache results, the transmission rate for caching problem~$i$ is
\begin{align}
R_i=\beta_i R_{\text{eq}}(N,i,\frac{M_i-M_{i+1}}{\beta_i})+\beta_i(K-i),\;i\in\mathcal{K}.\label{eq:existing1}
\end{align}
The first term on the right-hand side of~\eqref{eq:existing1} corresponds to the transmission rate for the first groups of users, and the second term corresponds to the transmission rate for the second group of users, which are without cache in problem~$i$. 

Therefore, by optimising the sum rate over the parameters $(\beta_1,\beta_2,\ldots,\beta_K)$, we achieve the following transmission rate
\begin{align}\label{Eq:existingwork1}
R_{\text{ex1}}(N,K,\mathbf{M})=\underset{(\beta_1,\ldots,\beta_K):\sum_{i=1}^{K}\beta_i=1}{\min}\sum_{i=1}^{K}R_i.
\end{align}

\subsubsection{Scheme~2~\cite{CentralizedUnequalCache2}}\label{Sec:ExistingScheme2} In this scheme, the problem of centralized caching with unequal cache sizes is formulated as an optimisation problem where it is assumed that the cache placement is uncoded, and the delivery phase uses linear coding. To characterize all possible uncoded placement policies, the parameter $a_{\mathcal{S}}$, $\mathcal{S}\subseteq\mathcal{K}$, is defined where $a_{\mathcal{S}}F$ represents the length of ${W}_{\ell,\mathcal{S}}$ as the fraction of $W_\ell$ stored in the cache of users in $\mathcal{S}$. Hence, these parameters must satisfy
\begin{align*}
\sum_{\mathcal{S}\subseteq\mathcal{K}} a_{\mathcal{S}}=1,
\end{align*}
and
\begin{align*}
\sum_{\mathcal{S}\subseteq\mathcal{K}:i\in\mathcal{S}} a_{\mathcal{S}}\leq\frac{M_i}{N},\;i\in\mathcal{K}.
\end{align*}

In the delivery phase, the server transmits
\begin{align*}
X_{\mathbf{d},\mathcal{T}}=\bigoplus_{j\in\mathcal{T}}W_{d_j}^{\mathcal{T}},
\end{align*}
to the users in $\mathcal{T}$ where $\mathcal{T}$ is a non-empty subset of $\mathcal{K}$. $W_{d_j}^{\mathcal{T}}$, which is a part of $W_{d_j}$, needs to be decoded at user~$j$, and cancelled by all the users in $\mathcal{T}\setminus\{j\}$. Therefore, $W_{d_j}^{\mathcal{T}}$ is constructed from subfiles  ${W}_{d_j,\mathcal{S}}$ where $\mathcal{T}\setminus\{j\}\subseteq \mathcal S$ and $j\notin \mathcal{S}$. To characterize all possible linear delivery policies, two sets of parameters are defined: (i)  $v_{\mathcal{T}}$ where $v_{\mathcal{T}}F$ represents the length of  $W_{d_j}^{\mathcal{T}},\;\forall j\in\mathcal{T}$, and consequently $X_{\mathbf{d},\mathcal{T}}$. (ii) $u_{\mathcal{S}}^{\mathcal{T}}$ where $u_{\mathcal{S}}^{\mathcal{T}}F$ is the length of $W_{d_j,\mathcal{S}}^{\mathcal{T}}$ which is the fraction of ${W}_{d_j,\mathcal{S}}$ used in the construction $W_{d_j}^{\mathcal{T}}$. In order to have a feasible delivery scheme, these parameters need to satisfy some conditions~\cite[equations (25)--(30)]{CentralizedUnequalCache2}. By considering $(\mathbf{a},\mathbf{u},\mathbf{v})$ as all the optimisation parameters, and $\mathcal{C}(N,K,\mathbf{M})$ as all the conditions that need to be met in the both placement and delivery phases, we achieve the following transmission rate
\begin{align}\label{Eq:existingwork2}
R_{\text{ex2}}(N,K,\mathbf{M})\hskip-2pt=\hskip-2pt\underset{\mathbf{d}}{\max}\hskip-2pt\left(\hskip-2pt\underset{(\mathbf{a},\mathbf{u},\mathbf{v}):\mathcal{C}(N,K,\mathbf{M})}{\min}\sum_{\mathcal{T}\in\mathcal{K}:\left|\mathcal{T}\right|\neq 0} v_{\mathcal{T}}\hskip-2pt\right).
\end{align}

%%-------------------------------------------------------------------------
%\begin{figure}[t]
%	\centering
%	\includegraphics[width=0.3\textwidth]{FiguresUnequalCacheSize/Example1.pdf}
%	\caption{Example 1} 
%	\label{Fig:Example1}
%\end{figure}
%%-------------------------------------------------------------------------

\section{Proposed Caching Scheme}
In this section, we first provide some insights into our proposed scheme using an example. We then propose a scheme for a system with two subgroups of users, one with a larger cache size than the other, i.e., $M_i=\hat{M}$, $1\leq i \leq L$, and $M_i={M}$, $L+1\leq i \leq K$, for some $\hat{M}>M$.

\subsection{An Example}
\begin{figure}[t]
	\centering
	\includegraphics[width=0.45\textwidth]{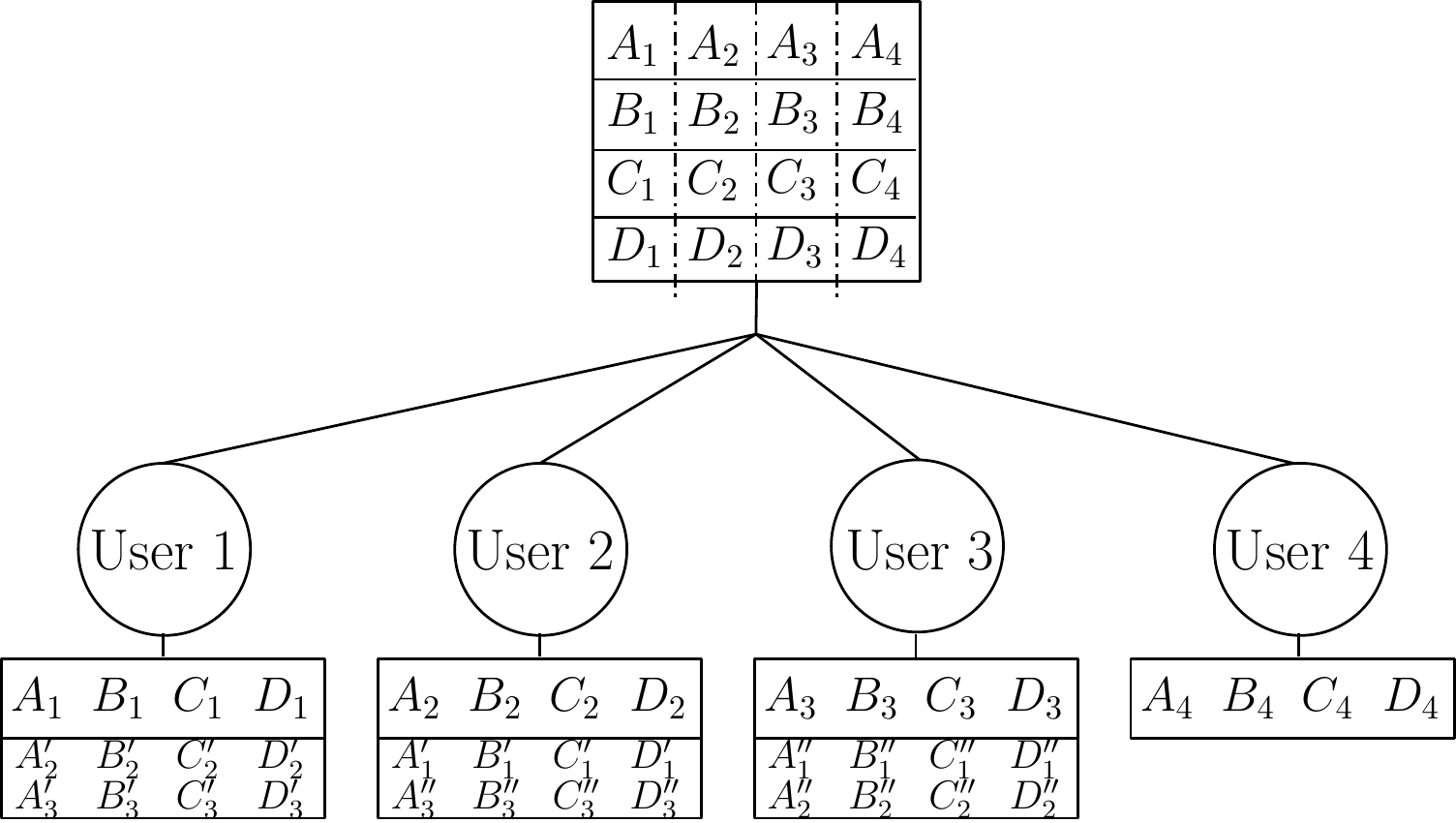}
	\vskip-10pt
	\caption{An example for our proposed scheme} 
	\vskip-12pt
	\label{Fig:AnExample}
\end{figure}

In our example, as shown in Fig.~\ref{Fig:AnExample}, we consider the case where the number of files in the server is four, denoted for simplicity by $(A,B,C,D)$, and the number of users is also four. The first three users have a cache of size $2F$ bits, and the forth one has a cache of size $F$ bits. First, we ignore the extra cache available at the first three users, and use the equal-cache scheme. This divides each file into four parts, and places $(A_i, B_i, C_i, D_i)$, $i\in\{1,2,3,4\}$, in the cache of user~$i$. Therefore, assuming without loss of generality that users~1,~2,~3 and~4 request $A$, $B$,  $C$ , and $D$ respectively, the server needs to transmit $A_2\oplus B_1$, $A_3\oplus C_1$, $B_3\oplus C_2$, $A_4\oplus D_1$, $B_4\oplus D_2$ and $C_4\oplus D_3$, and we achieve the rate of $R=3/2$ by ignoring the extra cache available at the first three users. Now, to utilize the extra cache available at users~1,~2, and~3, we look at what is going to be transmitted when ignoring these extra caches, and fill the extra caches to reduce the load of the transmission. In particular, we reduce the load of the transmissions which are only of benefit to the users with a larger cache size (i.e., $A_2\oplus B_1$, $A_3\oplus C_1$, $B_3\oplus C_2$). To do this, we divide $A_i$, $i\in\{1,2,3\}$ into two equal parts, $A'_i$ and $A''_i$. We do the same for $B_i$, $C_i$, and $D_i$, $i\in\{1,2,3\}$. We then place $(A'_2, B'_2, C'_2, D'_2)$ and $(A'_3, B'_3, C'_3, D'_3)$ in the extra cache of user~1, $(A'_1, B'_1, C'_1, D'_1)$ and $(A''_3, B''_3, C''_3, D''_3)$ in the extra cache of user~2, and $(A''_1, B''_1, C''_1, D''_1)$ and $(A''_2, B''_2, C''_2, D''_2)$ in the extra cache of user~3. Therefore, considering the extra cache available at the first three users, instead of $A_2\oplus B_1$, $A_3\oplus C_1$, $B_3\oplus C_2$, we just need to transmit $A''_2\oplus B''_1\oplus C'_1 $, and $A''_3\oplus B'_3\oplus C'_2$ to satisfy the demands of all users, and we achieve the rate $R=1$.

Note that what we did in the second part is equivalent to using the equal-cache scheme for a system with a server storing four files of size $\frac{3}{4}F$ bits, i.e., $A^*=(A_1,A_2,A_3)$, $B^*=(B_1,B_2,B_3)$, $C^*=(C_1,C_2,C_3)$, and $D^*=(D_1,D_2,D_3)$, and with three users each with a cache of size $2F$ bits. This can be seen by defining $A^*_{12}=(A'_1,A'_2)$, $A^*_{13}=(A''_1,A'_3)$, and $A^*_{23}=(A''_2,A''_3)$ for $A^*$, and also similarly for $B^*$, $C^*$, and $D^*$. Then we can check that $(A^*_\mathcal{T},B^*_\mathcal{T},C^*_\mathcal{T},D^*_\mathcal{T})$, $\mathcal{T}\in\{\{12\},\{13\},\{23\}\}$, is in the cache of user~$i$, $i\in\{1,2,3\}$ if $i\in\mathcal{T}$.

%-------------------------------------------------------------------------
%-------------------------------------------------------------------------

\subsection{Scheme with Two Levels of Caches}
In this subsection, we explain our proposed scheme for the system where the first $L$ users have a cache of size $\hat{M}F$ bits, and the last $K-L$ users have a cache of size $MF$ bits for some $M<\hat{M}$.

\subsubsection{An incremental placement approach}\label{Sec:IncreasingCache} We first describe a concept which is used later in our proposed scheme for the unequal-cache problem. Suppose that we initially have a system with $N$ files, and $K$ users each having a cache of size $MF$ bits. We use the equal-cache scheme described in Section~\ref{Sec:EqualCache} to fill the caches. 

We later increase the cache size of \textit{each} user by $(M'-M)F$ bits for some $M'>M$. The problem is that we are not allowed to change the content of the first $MF$ bits that we have already filled, but we want to fill the additional cache in such a way that the overall cache has the same content placement as the scheme described in Section~\ref{Sec:EqualCache} for the new system with $N$ files, and $K$ users each having a cache of size $M'F$ bits.

We present our solution when $M=\frac{tN}{K}$ and $M'=\frac{(t+1)N}{K}$ for some integer $t$. The solution can be easily extended to an arbitrary $M$ and $M'$. In the cache placement for the system with the parameters $(N,K,M)$, we divide $W_\ell$, $\ell\in\mathcal{N}$, into $\binom{K}{t}$ subfiles denoted by $W_{\ell,\mathcal{T}}$, and place the ones with $i\in\mathcal{T}$ in the cache of user~$i$. This means that we put $\binom{K-1}{t-1}$ subfiles of $W_\ell$ in the cache of each user. After increasing the cache of each user to $M'F$ bits, we further divide each subfile into $(K-t)$ parts denoted by $W_{\ell,\mathcal{T},j}$, $j\in\mathcal{K}\setminus\mathcal{T}$, and place  $W_{\ell,\mathcal{T},j}$ in the cache of user~$j$. This adds $W_{\ell,\mathcal{T},j}$, $j\notin\mathcal{T}$, to the cache of user~$j$ while keeping the existing content of the first $MF$ bits of user $j$, i.e., $W_{\ell,\mathcal{T},i}$ $j\in\mathcal{T}$, $i\in\mathcal{K}\setminus\mathcal{T}$. This means that we add
\begin{align*}
N\frac{\binom{K-1}{t}}{\binom{K}{t}(K-t)}F=\frac{N}{K}F=(M'-M)F\;\; \text{bits},
\end{align*}
to the cache of each user which satisfies the cache size constraint. Our cache placement for the system with the parameters $(N,K,M')$ becomes the same as the one described in Section~\ref{Sec:EqualCache} by merging all the parts $W_{\ell,\mathcal{T},j}$ which have the same $\mathcal{T}'=\mathcal{T}\cup\{j\}$ as a single subfile $W_{\ell,\mathcal{T}'}$, where $|\mathcal{T}'|=t+1$.

\subsubsection{Proposed Scheme} We here present our proposed scheme for the system where $M_i=\hat{M}$, $i\in\mathcal{L}$, $\mathcal{L}=\{1,2,\ldots,L\}$, and $M_i={M}$, $i\in\mathcal{K}\setminus\mathcal{L}$, for some $M<\hat{M}$.

Our placement phase is composed of two stages. In the first stage, we ignore the extra cache available at the first $L$ users, and use the equal-cache placement for the system with the parameters $(N,K,M)$. Hence, at the end of this stage, we can achieve the rate in~\eqref{Eq:Rate} by transmitting $X^{(\alpha)}_{\mathbf{d},\mathcal{S}_1}$, defined in~\eqref{Eq:Component1}, for any $\mathcal{S}_1\subseteq\mathcal{K}$ where $|\mathcal{S}_1|=t_\text{int}+1$, and $X^{(1-\alpha)}_{\mathbf{d},\mathcal{S}_2}$, defined in~\eqref{Eq:Component2}, for any $\mathcal{S}_2\subseteq\mathcal{K}$ where $|\mathcal{S}_2|=t_\text{int}+2$.

In the second stage of our placement phase, we fill the extra cache available at the first $L$ users by looking at what are going to be transmitted when ignoring these extra caches. To do so, we try to reduce the load of the transmissions which are intended only for the users with a larger cache size, i.e., $X^{(\alpha)}_{\mathbf{d},\mathcal{S}_1}$for any $\mathcal{S}_1\subseteq\mathcal{L}$ ($|\mathcal{S}_1|=t_\text{int}+1$), and $X^{(1-\alpha)}_{\mathbf{d},\mathcal{S}_2}$ for any $\mathcal{S}_2\subseteq\mathcal{L}$ ($|\mathcal{S}_2|=t_\text{int}+2$). These transmissions are constructed from the subfiles $W^{(\alpha)}_{\ell,\mathcal{T}_1}$, $\mathcal{T}_1\subseteq\mathcal{L}$, $|\mathcal{T}_1|=t_\text{int}$, and $W^{(1-\alpha)}_{\ell,\mathcal{T}_2}$, $\mathcal{T}_2\subseteq\mathcal{L}$, $|\mathcal{T}_2|=t_\text{int}+1$. These subfiles occupy
\begin{align}
\frac{\binom{L-1}{t_\text{int}-1}}{\binom{K}{t_\text{int}}}N\alpha F\hspace{-3pt}+\hspace{-3pt}\frac{\binom{L-1}{t_\text{int}}}{\binom{K}{t_\text{int}+1}}N(1-\alpha) F\;\; \text{bits},
\end{align}
of each user's cache, and the sum-length of these subfiles for any $\ell\in\mathcal{N}$ is
\begin{align*}
F'\triangleq \frac{\binom{L}{t_\text{int}}}{\binom{K}{t_\text{int}}}\alpha F+\frac{\binom{L}{t_\text{int}+1}}{\binom{K}{t_\text{int}+1}}(1-\alpha) F\;\;\text{bits}.
\end{align*}
Considering our aim in designing the second stage of our placement phase, we again use the equal-cache placement for the subfiles $W^{(\alpha)}_{\ell,\mathcal{T}_1}$, $\mathcal{T}_1\subseteq\mathcal{L}$, $|\mathcal{T}_1|=t_\text{int}$, and $W^{(1-\alpha)}_{\ell,\mathcal{T}_2}$, $\mathcal{T}_2\subseteq\mathcal{L}$ $|\mathcal{T}_2|=t_\text{int}+1$ while considering the extra cache available at the first $L$ users. 
This means that we use the equal-cache scheme for a system with $N$ files of size $F'$ bits, and $L$ users each having a cache of size $M'F'$ bits where
\begin{align}\label{Eq:CacheSize2}
M'\hspace{-2pt}F'\triangleq\hspace{-3pt}\frac{\binom{L-1}{t_\text{int}-1}}{\binom{K}{t_\text{int}}}N\alpha F\hspace{-3pt}+\hspace{-3pt}\frac{\binom{L-1}{t_\text{int}}}{\binom{K}{t_\text{int}+1}}N(1-\alpha) F\hspace{-3pt}+\hspace{-3pt}(\hat{M}-M){F}.
\end{align}
Note that we are not allowed to change what we have already placed in the cache of the first $L$ users in the first stage. Otherwise, we cannot assume that, from the delivery phase when ignoring the extra caches, the transmissions $X^{(\alpha)}_{\mathbf{d},\mathcal{S}_1}$ where $\mathcal{S}_1=\mathcal{T}_1\cup\{j\}$, $|\mathcal{T}_1|=t_\text{int}$, $\mathcal{T}_1\subseteq \mathcal{L}$, $j\in\mathcal{K}\setminus\mathcal{L}$, and $X^{(1-\alpha)}_{\mathbf{d},\mathcal{S}_2}$ where $\mathcal{S}_2=\mathcal{T}_2\cup\{j\}$, $|\mathcal{T}_2|=t_\text{int}+1$, $\mathcal{T}_2\subseteq \mathcal{L}$, $j\in\mathcal{K}\setminus\mathcal{L}$, can still be decoded by target users. Therefore, we employ our proposed solution in Section~\ref{Sec:IncreasingCache} for using the equal-cache scheme for the second time. 

Two scenarios can happen in the second stage.

\textit{Scenario~$1$} where $M'\leq N$: In this scenario, we achieve the rate
\begin{align*}
R_\text{ueq}(N,K,L,\hat{M},M)\hspace{-3pt}=\hspace{-3pt}R_\text{eq}(N,K,M)\hspace{-3pt}-\hspace{-3pt}R'\hspace{-3pt}+\hspace{-3pt}R_{\text{eq}}(N,L,M')\frac{F'}{F},
\end{align*}
where 
\begin{align*}
R'= \alpha \frac{\binom{L}{t_\text{int}+1}}{\binom{K}{t_\text{int}}}+(1-\alpha)\frac{\binom{L}{t_\text{int}+2}}{\binom{K}{t_\text{int}+1}}.
\end{align*}
$R'F$ is the load of the transmissions intended only for the users with a larger cache size if we ignore their extra caches (or equivalently if we just utilize the first stage of our placement phase). $R_\text{eq}(N,L,M')F'$ is the new load of the transmissions intended only for the users with a larger cache size at the end of the second stage. 

\textit{Scenario~$2$} where $M'> N$: In this scenario, we also use memory sharing between the case with $\hat{M}=\Phi$, where 
\begin{align*}
\Phi\triangleq M-\frac{\binom{L-1}{t_\text{int}-1}}{\binom{K}{t_\text{int}}}N\alpha-\frac{\binom{L-1}{t_\text{int}}}{\binom{K}{t_\text{int}+1}}N(1-\alpha)+N\frac{F'}{F},
\end{align*}
and the case with $\hat{M}=N$. In the system with $\hat{M}=\Phi$, according to~\eqref{Eq:CacheSize2}, we have $M'=N$, and we achieve the rate $R_\text{eq}(N,K,M)-R'$. In the system with $\hat{M}=N$, we can simply just remove the first $L$ users as they can cache the whole files in the server, and we achieve the rate $R_{\text{eq}}(N,K-L,M)$. Therefore, in this scenario, we achieve the rate
\begin{align*}
R_\text{ueq}(N,K,L,\hat{M},M)=&\gamma (R_\text{eq}(N,K,M)-R')\\
&\hskip25pt+(1-\gamma)R_{\text{eq}}(N,K-L,M),
\end{align*}
where $0\leq\gamma\leq1$, and is calculated using $\hat{M}=\gamma \Phi+(1-\gamma)N$.

\section{Comparison with existing works}
\begin{figure}[t]
	\centering
	\includegraphics[width=0.45\textwidth]{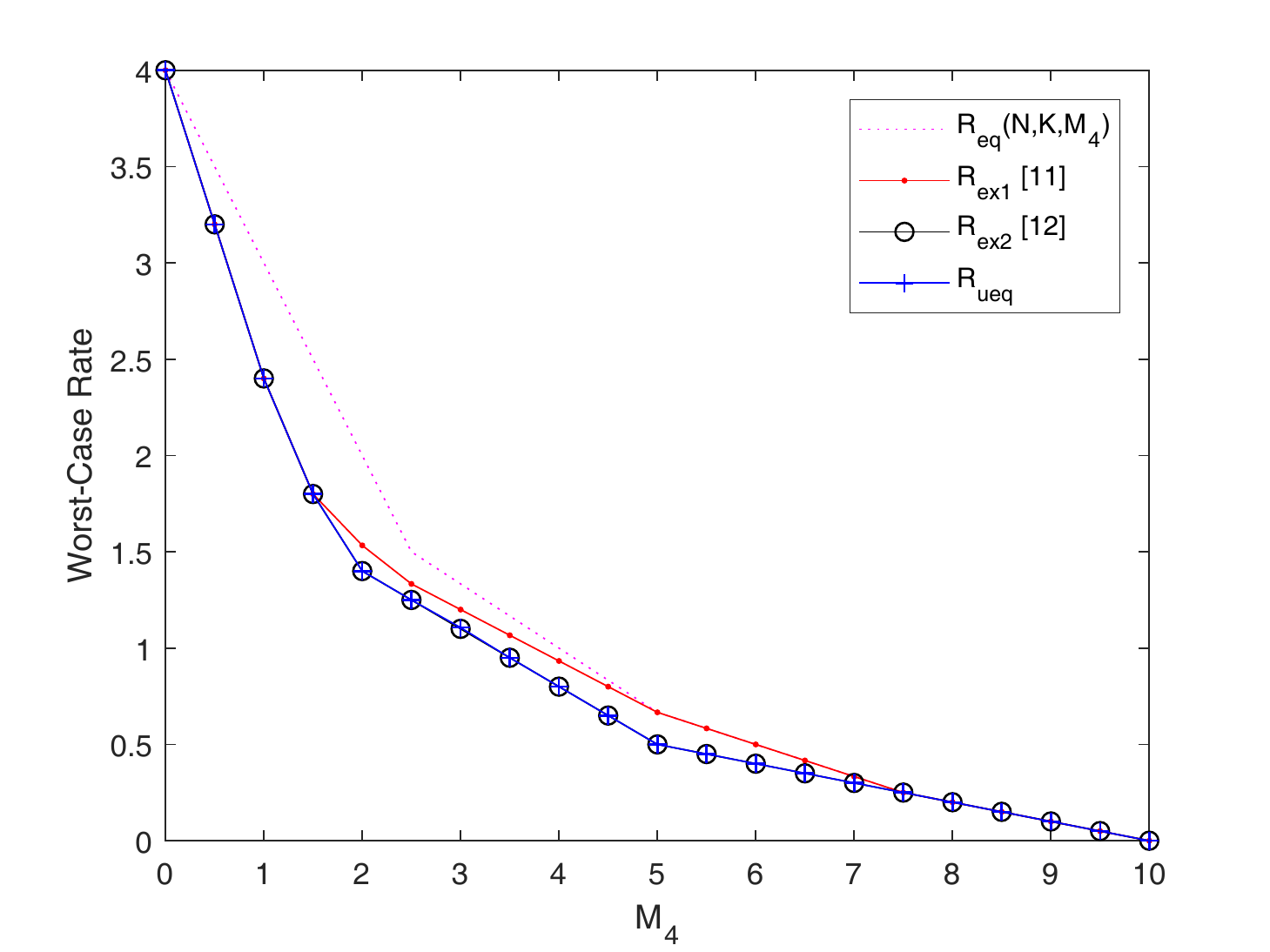}
	\vskip-15pt
	\caption{Comparing the worst-case transmission rate of the proposed scheme with the existing ones for the system with $N=10$, $K=4$, $M_1=M_2=3M_3=3M_4$.} 
	\vskip-15pt
	\label{Fig:Comparison}
\end{figure}

In this section, we present our numerical results comparing our proposed scheme with the existing works, described in Section~\ref{Sec:ExistingWorks}. Our numerical results, characterizing the trade-off between the worst-case transmission rate and cache size for systems with two levels of cache sizes, suggest that our scheme outperforms the scheme by Saeedi Bidokhti et al.~\cite{CentralizedUnequalCache1}. 
Considering the work by Ibrahim et al.~\cite{CentralizedUnequalCache2}, as the complexity of the solution grows exponentially with the number of users, we implemented that work for systems with up to four users. Our numerical evaluations suggest that our scheme performs withing a multiplicative factor of 1.11 from that scheme, i.e., $1\leq\frac{R_\text{ueq}}{R_{\text{ex2}}}\leq1.11$. As an example, this comparison is shown in Fig.~\ref{Fig:Comparison} for a four-user system with the parameters $N=10$, $K=4$, $M_1=M_2=3M_3=3M_4$. For these parameters, our scheme performs as well as the work by Ibrahim et al.~\cite{CentralizedUnequalCache2} without needing to solve an optimisation problem to obtain the scheme.

%-------------------------------------------------------------------------
%-------------------------------------------------------------------------

\vspace{-5pt}
\section{Conclusion}
We addressed the problem of centralized caching with unequal cache sizes. We proposed an explicit scheme for the system with a server of files connected through a shared error-free link to a group of users where one subgroup is equipped with a larger cache size than the other. Numerical results comparing our scheme with existing works showed that our scheme improves upon the existing explicit scheme by having a lower worst-case transmission rate over the shared link. Numerical results also showed that our scheme achieves within a multiplicative factor of 1.11 from the optimal worst-case transmission rate for schemes with uncoded placement and linear coded delivery without needing to solve a complex optimisation problem.

\vspace{-5pt}
\bibliographystyle{IEEEtran}

\end{document}